\documentclass[fleqn,conference]{IEEEtran}
\pdfoutput=1 
\makeatletter
\def\ps@headings{%
\def\@oddhead{\mbox{}\scriptsize\rightmark \hfil \thepage}%
\def\@evenhead{\scriptsize\thepage \hfil \leftmark\mbox{}}%
\def\@oddfoot{}%
\def\@evenfoot{}}
\makeatother
\usepackage{tabularx}
\usepackage{amssymb}
\usepackage{graphicx}
\usepackage{tikz}
\usepackage{amsmath}
\usepackage{mathtools}
\usepackage{subcaption}
\usepackage{pgfplots}
\usepackage{color, colortbl}
\usepackage{xcolor}
\usepackage[utf8]{inputenc}
\usepackage{tabularx,ragged2e,booktabs}
\usepackage{makecell}
\usetikzlibrary{matrix,positioning}

\usepgfplotslibrary{groupplots}

\newcolumntype{C}{>{\Centering\arraybackslash}X} 
\definecolor{Gray}{gray}{0.9}
\pgfplotsset{width=\linewidth,compat=1.14,
	    every axis/.append style={
		label style={font=\small},
		tick label style={font=\small},
		enlarge x limits={abs=1}},
 		 /pgfplots/ybar legend/.style={
		/pgfplots/legend image code/.code={%
		\draw[##1,/tikz/.cd,yshift=-0.25em]
		(0cm,0cm) rectangle (3pt,0.8em);},
},
}
\newcommand{\squeezeup}{\vspace{-4.0mm}}

\begin{document}
	
\title{Proactive Allocation as Defense for Malicious Co-residency in Sliced 5G Core Networks}
\author{\IEEEauthorblockN{Danish Sattar and Ashraf Matrawy\\}
	\IEEEauthorblockA{
		Carleton University,
		Ottawa, ON Canada\\
		Email: \{Danish.Sattar, Ashraf.Matrawy\}@carleton.ca}}
\maketitle

\begin{abstract}
Malicious co-residency in virtualized networks poses a real threat. The next-generation mobile networks heavily rely on virtualized infrastructure, and network slicing has emerged as a key enabler to support different virtualized services and applications in the 5G network. However, allocating network slices efficiently while providing a minimum guaranteed level of service as well as providing defense against the threat of malicious co-residency in a mobile core is challenging. To address this question, in our previous work, we proposed an optimization model to allocate slices. In this work, we analyze the defense against the malicious co-residency using our optimization-based allocation. 
\end{abstract}

\IEEEpeerreviewmaketitle

\section{Introduction}
Network Slicing has been proposed to cope with the ever-growing demand for flexibility and scalability in 5G mobile network \cite{NGNMWP,7926920}. The recent advancements in Network Function Virtualization (NFV) have enabled next-generation mobile networks to employ concepts like network slicing to satisfy diverse requirements from various new applications \cite{8685766}. The Next Generation Mobile Network Alliance (NGMN) defined network slicing as running multiple services with different requirements such as performance, security, availability, reliability, mobility, and cost as an independent logical network on the shared physical infrastructure \cite{NGNMWP,8039298}. An end-to-end slice is created by pairing the RAN and core network slice, but the relationship between both slices could be 1-to-1 or 1-to-M ~\cite{DBLP:journals/corr/LiWPU16,5gamericas}. 

One of the key requirements for network slicing is resource isolation between different slices \cite{8685766,NGNMWP}. However, guaranteeing resource isolation between slices that share the common physical infrastructure is challenging \cite{8039298}. The sharing of common physical resources between slices could lead to information leakage and side-channel attacks \cite{8056951,180210}. The side-channel attacks can be used to determine co-residency and extract valuable information (e.g., cryptographic keys \cite{191010d}) from the victim slices or perform Denial-of-Service attacks \cite{secureslicingCNS2019}. There are several types of side-channel attacks that can be used to determine co-residency by using different shared resources such as CPU cache, main memory, and network traffic \cite{191010d}. Therefore, it is paramount to provide defense against malicious co-residency and minimize side-channel attacks. 


In this paper, our focus is to minimize the success rate of getting a malicious co-residency with a victim slice. We analyze the impact of optimization-based slice allocation on malicious co-residency. We also discuss additional measures that can be taken to further minimize the risk of co-residency as well as some reactive measures. 
%
%

The rest of this paper is organized as follows. In Section~\ref{sec:COR:relatedwork}, we present the literature review on side-channel attacks and co-residency. The threat model is presented in section  \ref{sec:COR:Tmodel}. Section~\ref{sec:COR:mathmodel} provides an overview of the optimization model for 5G network slicing. We discuss our simulation setup in section \ref{sec:COR:psimulation}. In the section~\ref{sec:COR:RndD}, we discuss our results and lastly, section~\ref{sec:COR:con}, we present our conclusion.

\section{Related Work}
\label{sec:COR:relatedwork}
The next-generation mobile networks will have similar co-residency issues as cloud networks since both networks share some properties, i.e., shared resources and multi-tenancy. The work on the co-residency issue in the 5G network is currently limited due to its infancy. Therefore in this section, we describe some of the state-of-the-artwork on co-residency detection in could networks, which would still be applicable in the 5G network because both networks share virtualized infrastructure. 

Network traffic is one of the shared resources that can be used to determine co-residency. In \cite{10.1145/2381913.2381915}, A. Bates \textit{et al.} used network traffic watermarking to detect co-residency with the victim. In the proposed scheme, the attacker launches multiple VM instances called FLOODER that communicate with the CLIENT, which is outside the cloud network. The CLIENT sends legitimate traffic to the target (victim) server that resides inside the cloud network. The FLOODER VMs flood the network with traffic to cause network delays, and the CLIENT analyzes these delays to determine which FLOODER is co-resided with the target server. 
Another aspect of network traffic can be exploited by analyzing Round Trip Time (RTT) to detect co-residency. Such a method is discussed by A. Atya \textit{et al.} in \cite{8056951}. In the proposed work, TCP handshake is used to measure the RTT (in some cases, ICMP was also be used) to determine co-residency. RTT is calculated from multiple sources and vantage points to increase the accuracy of detecting co-residency. A migration scheme was proposed to defend against co-residency attacks. An extension of their work is also discussed in \cite{8648158}.   

CPU cache-based side-channel attacks are commonly used to detect co-residency with the victim VM. 
Authors Y. Zhang {et al.} \cite{5958037} used L2 cache to detect co-residency in the cloud environment. The objective of their works was to use side-channel to detect undesired co-residency. The basic idea of HomeAlone was to coordinate with other friendly VMs and analyze the cache usage to determine if there are any undesired VMs hosted on the same hypervisor. 



\section{Threat Model}
\label{sec:COR:Tmodel}
\textbf{Assumptions:}
Our threat model assumes that network slicing is supported by the target network, the Evolved Packet Core (EPC) supports migration of the slice components, the slice operator or users can migrate the slice(s), on the servers, multi-tenancy is supported by the slice operator, adversaries do not know about the operators' allocation scheme, and adversaries can successfully determine the co-residency with the victim slice.

\textbf{Adversary:} The adversaries can launch multiple VMs and check for the co-residency with the victim. If co-residency is found then, the attacker could launch the next attack(s). If not, remove the slice and launch new ones and repeat the process. We assume that adversaries are not colluding. 

\section{Optimization Model}
\label{sec:COR:mathmodel}
In our previous work \cite{secureslicingCNS2019}, we proposed an optimization model to mitigate DDoS attacks. The proposed model mitigated DDoS attacks using intra-slice (between slice components) and inter-slice\footnote{Please note that in this paper we did not consider the inter-slice isolation} (between slices) isolation. In addition to providing defense against DDoS, it can optimally allocate slices. Our model allocated slices to the least loaded\footnote{All servers have same max. CPU capacity so least loaded is also least utilized} servers and finds the minimum delay path (Eq. \ref{eq1}). The optimization model also fulfills several requirements of the 5G network. It can guarantee the end-to-end delay and provide different levels of slice isolation for reliability and availability as well as it assures that allocation does not exceed the available system resources. In our model, we only considered CPU, bandwidth, VNF processing delay, and link delay. Intra-Slice isolation could increase the availability of a slice. If all components of the slice are hosted on the same hypervisor, any malfunction could result in the slice unavailability. However, different levels of intra-slice isolation can ensure that full or partial slice remains available.  

A summary of the optimization model presented in  \cite{secureslicingCNS2019} is provided here for better readability. More details can be found in \cite{secureslicingCNS2019}. We use an undirected graph $G_p=(N_p,L_p)$ to represent the physical 5G core network topology. All the nodes in the network (i.e. servers, switches, routers and other devices present in the network) are represented by $N_p$, and $L_p$ denotes all the physical links between the nodes. A slice request is denoted by a directed graph $G_v = (N_v,L_v)$, where $N_v=(N_c\cup N_d)$ contains all the slice virtual network functions, the control and data plane virtual functions are represented by $N_c$ and $N_d$, respectively and $L_v$ represents requested links. Each edge in the directed graph is represented by $(i,j) \in L_v$. Each slice request is associated with end-to-end delay ($d_{E2E}$), intra-slice isolation ($K^c_{rel},K^d_{rel}$), and each VNF in a slice is associated with a computing demand ($R^i$), and bandwidth (BW) requirement between VNF \textit{i} and VNF \textit{j} ($R^{ij}$). The description of all variable is provided in table~\ref{tab:parameters}.
\begin{table}[htbp]
	\centering
	\captionof{table}{Variable Description} \label{tab:parameters}
	\begin{tabular}{|r|l|}
		\hline
		\textbf{Parameter} & \textbf{Description} \\ \hline
        $N_p$ & Set of physical Nodes \\
		$L_p$ & Physical links between nodes \\
		$\sigma_k$ & Current CPU allocation of physical node $k$ \\
		$\sigma_{ef} $ & \makecell{Current BW allocation of physical link\\between nodes $e,f$}\\
		$\sigma_k^{max}$ & Maximum CPU capacity of physical node $k$ \\
		$\sigma^{max}_{ef} $ & \makecell{Maximum BW capacity of physical link\\between nodes $e,f$}\\
		$T_{ef}$ & Physical link delay between node $e,f$ \\
		$\Delta_k$ & Physical node $k$ processing delay \\
		$\Delta^i$ & VNF $i$ processing delay \\
		$N_c$ & Requested set of slice control plane functions \\
		$N_d$ & Requested set of slice data plane functions \\
		$N_v$ & Requested set of slice VNFs ($N_c\cup N_d$) \\
		$L_v$ &  Requested virtual links of a slice\\
		$R^i$ & Requested CPU resource by a VNF $i$ \\
		$R^{ij}$ & Requested BW resource between VNF $i,j$\\
		$d_{E2E}$ & Requested End-to-End delay\\
		$K^c_{rel}$ & Requested intra-slice isolation for Control Plane\\
		$K^d_{rel}$ & Requested intra-slice isolation for Data Plane\\
		$u^i_k$ & Indicates the assignment of VNF $i$ to EPC node $k$ \\
		$y^{ij}_{ef}$ & \makecell {Indicates the assignment of link $(e,f)$\\ for VNF graph edge $(i,j)$} \\
		\hline
	\end{tabular} \par
	\bigskip
	\squeezeup
	\squeezeup
\end{table}
\begin{equation}\label{eq1}
\begin{multlined}
\emph{Minimize}\\
\sum _{i\in{N}_{v}}\sum _{k\in {N}_{p}} \left(\sigma_k+ R^i\right)u^i_k
+\sum_{(i,j)\in{L}_{v}}\sum _{\substack{(e,f)\in {L}_{p} \\(e\neq f)}}T_{ef} y^{ij}_{ef}
\end{multlined}
\end{equation}
The objective function of our optimization model (eq. \ref{eq1}) allocates the slice to least loaded physical nodes and find minimum delay path. The first term will assign the slice request to the least loaded servers. The second term will find the minimum delay path.

The objective function is subjected to several Mixed-Integer Linear Programming (MILP) constraints. 
\begin{enumerate}
	\item \textbf{Slice Assignment, Placement and Resource Budget}
	\begin{equation}\label{eq2}
	\begin{multlined}
	\sum _{k \in N_p} u^i_k=1 \hspace{.5cm}\forall i \in N_v
	\end{multlined}
	\end{equation}
	\begin{equation}\label{eq3}
	\begin{multlined}
	\sum _{i \in N_v} \left(R^i + \sigma_k\right) u^i_k \leq \sigma^{max}_k \hspace{0.5cm}\forall k \in N_p
	\end{multlined}
	\end{equation}
	\begin{equation}\label{eq4}
	\begin{multlined}
	\sum _{(i,j) \in L_v} (R^{ij}+\sigma^{ij}_{ef}) y^{ij}_{ef} \leq \sigma^{max}_{ef} \hspace{0.5cm} \forall (e,f) \in L_p
	\end{multlined}
	\end{equation}
	\begin{equation}\label{eq5}
	\begin{multlined}
	\sum _{i \in N_v} R^i \leq \sum_{k \in N_p} \left(\sigma_k^{max} - \sigma_k\right)
	\end{multlined}
	\end{equation}
	\begin{equation}\label{eq6}
	\begin{multlined}
	\sum _{\substack{f\in {N}_{p} \\(e\neq f)}}\left(y^{ij}_{ef}-y^{ij}_{fe}\right)=\left( u^i_e-u^j_e\right)\\
	\hspace{0.5cm} i\neq j, \forall (i,j) \in L_v, \forall e \in N_p
	\end{multlined}
	\end{equation}
	
	\begin{equation}\label{eq7}
	\begin{multlined}
	u_k^i \in \{0,1\} \hspace{.5cm} \forall k \in N_p,  \forall i \in N_v
	\end{multlined}
	\end{equation}
	\begin{equation}\label{eq8}
	\begin{multlined}
	y_{ef}^{ij} \ge 0 \hspace{.5cm} \forall (i,j) \in L_v, \forall (e,f) \in L_p
	\end{multlined}
	\end{equation}
	The constraint (\ref{eq2}) ensures that each VNF is assigned to an exactly one server. Constraints (\ref{eq3}) and (\ref{eq4}) guarantee that allocated VNF resources do not exceed the physical servers' processing capacity and link bandwidth, respectively. A slice CPU demand should not exceed the remaining CPU capacity of the entire system. This is ensured by constraint (\ref{eq5}) since partial allocation of a slice is not the desired behavior. The conservation of flows, i.e., the sum of all incoming and outgoing traffic in the physical nodes that do not host VNFs should be zero is enforced by the constraint (\ref{eq6}), and this constrains also ensures that there is a path between VNFs. Constraints (\ref{eq7}) and (\ref{eq8}) ensures that $u_k^i$ and $y_{ef}^{ij}$ are binary and integer, respectively.
	\item \textbf{End-to-End Delay:}
	\begin{equation}\label{eq12}
	\begin{multlined}
	\sum _{\substack{(i,j)\in {L}_{v} }} \sum _{\substack{(e,f)\in {L}_{p}\\e\neq f}}T_{ef}y^{ij}_{ef} +  \sum_{i\in N_v} \left(\Delta^i +\sum_{k\in N_p} \Delta_k u_k^i \right)\leq d_ {E2E}
	\end{multlined}
	\end{equation}
	\begin{equation}\label{eq12a}
	\begin{multlined}
	T_{ef}=\frac{\sigma_{ef}}{\sigma^{max}_{ef}} \delta  + T_{ef,init} \hspace{.3cm} \forall (e,f) \in L_p
	\end{multlined}
	\end{equation}
	Constraint (\ref{eq12}) guarantees end-to-end delay for a slice in the current network state. End-to-end delay includes link delay, VNF processing delay, and physical node processing delay. Each time when a virtual link $(i,j)\in L_v$ is assigned to a physical link $(e,f)\in L_p$, it increases $T_{ef}$. $T_{ef}$ is a function of link utilization, and it is calculated using eq. (\ref{eq12a}), where $T_{ef,init}$ is the initial delay assigned to the link $(e,f)\in L_p$ and $\delta$ is the maximum increase in delay.
	
	\item \textbf{Intra-Slice Isolation}
	\begin{subequations}\label{eq10}
		\begin{align}
		\sum _{i \in N_c} u^i_k\leq K_{rel}^c \hspace{.3cm}\forall k \in N_p, K_{rel}^c=1,2,3...\\
		\sum _{i \in N_d} u^i_k\leq K_{rel}^d \hspace{.3cm}\forall k \in N_p, K_{rel}^d=1,2,3...
		\end{align}
	\end{subequations}
	It might be required to have different levels of intra-slice isolation for control and data plane. In constraints (\ref{eq10}), $K_{rel}^c$ and $K_{rel}^d$  ensure the intra-slice isolation for control plane and data plane, respectively. Intra-slice isolation can improve the availability of a slice. 
\end{enumerate}
\section{Simulation Setup}
\label{sec:COR:psimulation}
\begin{figure}[!ht]
	\centering
	\includegraphics[width=\linewidth,keepaspectratio=true]{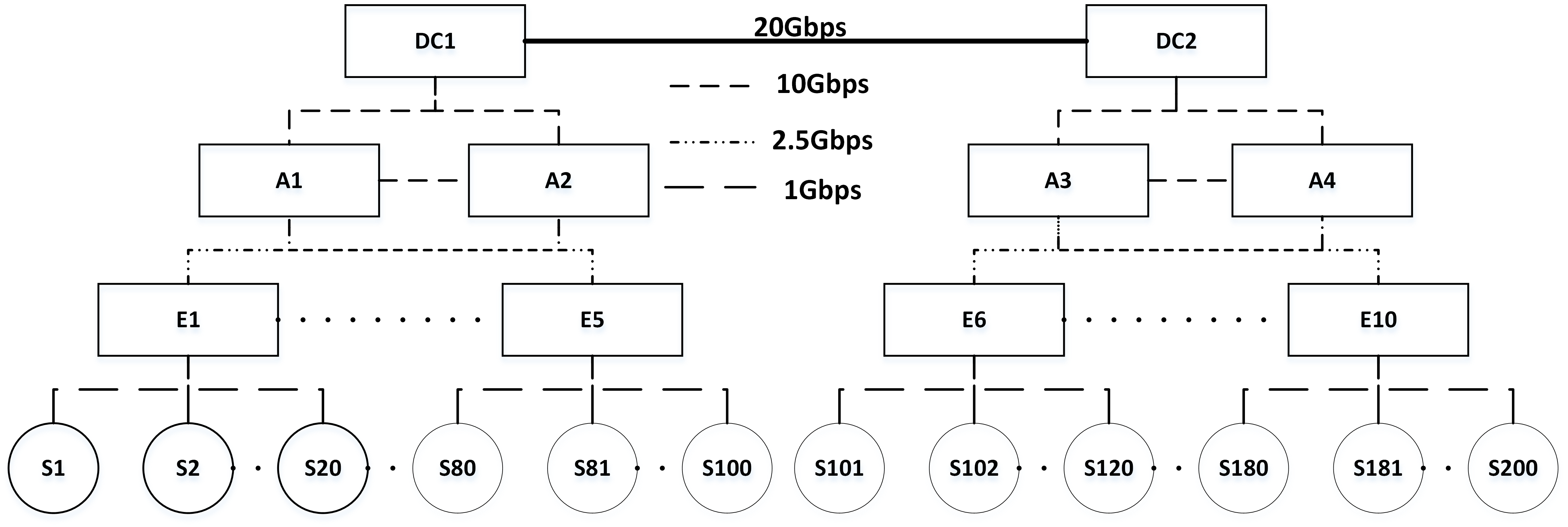}
	\caption{Simulation Topology}
	\label{fig:Simtopology}
	\squeezeup
\end{figure}

MATLAB is used to calculate the optimization solution and perform pre/post-processing of data. The preprocessing of data involves reading current network topology, slice requests, and updating the optimization model. In the post-processing, we update the network topology after a slice is allocated. AMPL is used to model optimization algorithm, and CPLEX 12.9.0.0 is used as a MILP solver. The optimization model is evaluated on Intel Core i7-8700 3.2 GHz with 32 GB RAM. We simulate 200 servers as shown in Fig.~\ref{fig:Simtopology} (we used similar topology to \cite{8676260}). Other parameters used for the evaluation are listed in Table~\ref{tab:Simparameters}. In our simulation, we vary the level of intra-slice isolation using the $K_{rel}$ parameter. This parameter provides the upper limit for how many VNFs can be placed on one physical server. For $K_{rel}$ 1 to 10, all slices request the same level of intra-slice isolation. The average overall CPU utilization of the entire system is also restricted at 50\%, 75\%, 80\%, 85\%, 90\%, and 95\% ($\pm \Delta 0.5\%$). For each average CPU utilization (ACU), we vary the $K_{rel}$ (e.g., at average CPU utilization 50\%, $K_{rel}$ will be varies from 1 to 10). 
\begin{table}[ht]
	\centering
	\captionof{table}{Simulation parameters} \label{tab:Simparameters}
	\begin{tabular}{r||l}\toprule[1.5pt]
		
		\bf Parameter		 & \bf Value 	\\\midrule
		\rowcolor{Gray}
		$N_p$				 &  200			\\
		$\sigma_k^{max}$	 &	25 GHz	\\
		\rowcolor{Gray}
		$K_{rel}$ 			 &	1-10		     \\
		$N_v$				 &	10			  \\
		\rowcolor{Gray}
		$R^{ij}$             &  40-60 Mbps	   \\
		$R^i$                &  0.55-1.6 GHz	\\
		\rowcolor{Gray}
		$\Delta^i$           &  0.2-0.6 ms	      \\
		$\Delta_k$           &  0.2 ms	        \\
		\rowcolor{Gray}
		$\delta$             &  3.5 ms	\\
		$T_{ef},init$        &  0.13 ms	\\
		\rowcolor{Gray}
		Total Attacker Slice Requests & 	500 		\\
		Target Slice &	50 \\
		\bottomrule[1.25pt]
		\end {tabular}\par
		\bigskip
		\squeezeup
	\end{table}

In each simulation, the attacker requests allocation of a slice and determines if there is a co-residency with the victim slice (i.e., one or more VNFs of the victim slice are allocated on the same hypervisor as the attacker). If co-residency is found, it is considered as a success (we assume that attacker will move to the next step of their objective in the real world), and if no co-residency is found then, it is considered as failure. In either case, we deallocate the attacker slice and request a new slice. We repeat this process 500 times and calculate the average success rate.  \textbf{The success is defined as if one or more VNFs of a victim slice are allocated on the same hypervisor as the attacker}. We only generate 500 attacker requests once at the beginning of the simulation. To simulate a more realistic scenario, every 60 seconds a legitimate slice is deallocated, and a new slice is allocated. The target slice is never deallocated. 
%

\section{Results and Discussion}
\label{sec:COR:RndD}

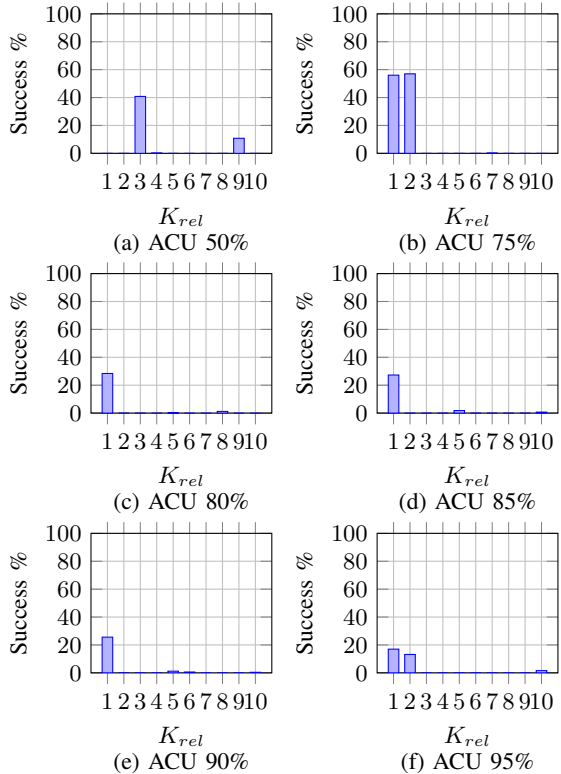
\begin{figure}[h]
	\centering
	
	\begin{center}
		\begin{tikzpicture}
		
		\begin{groupplot}[
		group style={
			group name=my plots,
			group size=2 by 3,
			horizontal sep=1.4cm,
			vertical sep=1.6cm,
		},
		ybar=0pt,
		xtick=data,
		ylabel = {Success \%},
		xlabel={$K_{rel}$},
		/pgf/bar width=4pt,
		grid=major,
		ymax=100,
		ymin=0,
		width=0.45\linewidth
		]
		\nextgroupplot[legend to name={CommonLegend},legend style={legend columns=5}]
		\addplot+  table[x=K,y=50] {Data/S1_M2_Success.txt};
		\nextgroupplot
		\addplot+  table[x=K,y=75] {Data/S1_M2_Success.txt};
		\nextgroupplot
		\addplot+  table[x=K,y=80] {Data/S1_M2_Success.txt};
		\nextgroupplot
		\addplot+  table[x=K,y=85] {Data/S1_M2_Success.txt};
		\nextgroupplot
		\addplot+  table[x=K,y=90] {Data/S1_M2_Success.txt};
		\nextgroupplot
		\addplot+  table[x=K,y=95] {Data/S1_M2_Success.txt};
		\end{groupplot}
		\path (my plots c1r1.south east) -- node[yshift=-21mm]{\ref{CommonLegend}} (my plots c2r1.south west);
		\node[text width=6cm,align=center,anchor=north] at ([yshift=-7mm]my plots c1r1.south) {\subcaption{ACU 50\%}\label{subplot:50}};
		\node[text width=6cm,align=center,anchor=north] at ([yshift=-7mm]my plots c2r1.south) {\subcaption{ACU 75\%\label{subplot:75}}};
		\node[text width=6cm,align=center,anchor=north] at ([yshift=-7mm]my plots c1r2.south) {\subcaption{ACU 80\%\label{subplot:80}}};
		\node[text width=6cm,align=center,anchor=north] at ([yshift=-7mm]my plots c2r2.south) {\subcaption{ACU 85\%\label{subplot:85}}};
		\node[text width=6cm,align=center,anchor=north] at ([yshift=-7mm]my plots c1r3.south) {\subcaption{ACU 90\%\label{subplot:90}}};
		\node[text width=6cm,align=center,anchor=north] at ([yshift=-7mm]my plots c2r3.south) {\subcaption{ACU 95\%\label{subplot:95}}};
		\end{tikzpicture} 
	\end{center}
\squeezeup
	\caption{Comparison of Co-Residency Success rate for different average CPU utilization}\label{fig:comparisonCPU}
	\squeezeup
\end{figure}
Fig. \ref{fig:comparisonCPU} shows the relationship between different levels of intra-slice isolation and the success rate of getting a co-residency with any network function of the target slice. In the figure, there is a relatively higher chance of getting co-residency when $K_{rel}\leq3$ and lower ACU because of two reasons. First, at $K_{rel}\leq3$, the network functions are more spread across the network, which increases the chances of getting co-residency with a specific target slice. Second, at relatively lower ACU\footnote{$ACU<50\%$ does not yield meaningful results due to low resource utilization. Therefore, we have not shown those results here } there are more opportunities to get co-allocation. Whereas at higher $K_{rel}\ge4$ and $ACU\ge80\%$, we see a significant decrease in the success rate of getting a co-residency. For instance, $K_{rel}=1$ and $ACU=75\%$, the success rate is 56\% whereas $K_{rel}=1$ and $ACU=80\%$, the success rate is only 29\% (almost 50\% reduction in success rate). At $K_{rel}\ge4$, the slice would have a certain degree of isolation as well as at $ACU>75$ present a more realistic scenario for the slice operator because the network resources will be better utilized. Please note that a detailed analysis of the optimization model's performance and efficiency is presented in \cite{secureslicingCNS2019, 8676260}.
\section{Defense}
There are few methods that can be employed to reduce the threat of malicious co-residency.  
\begin{itemize}
	\item Migrate the target slice to a different location (hypervisor). It could be a slice operator or the user (if allowed) that can migrate the slice.  
	\item Detect anomalous behavior for slice allocation requests (e.g., monitor IP/MAC/unique user ID or some other parameters) and take necessary preventive measures.
	\item Limit the number of slice requests allowed perusers within a specified period and the total number of requests.
	\item Randomize the time between slice requests and creation, thus making it harder to infer the allocation scheme. 
\end{itemize}
\section{Conclusion}
\label{sec:COR:con}
In this paper, we presented an analysis of the impact of optimization-based slice allocation on malicious co-residency. Our optimization model inherently provides a proactive defense against malicious co-residency. The success rate of getting co-residency with the target slice decreases with the increase in $K_{rel}$ levels and Average CPU Utilization of the system. The selection of $K_{rel}$ depends on several factors, i.e., cost, security, and performance. For instance, if a slice requires higher level of DDoS protection then lower $K_{rel}$ might be required to provide high availability (e.g., $K_{rel} = 1$). Whereas to reduce the cost and the risk of malicious co-residency, a slice might require higher $K_{rel}$ (e.g., $K_{rel}\ge4$). Another factor that could impact the selection of $K_{rel}$ is the state of the slice operator's network\footnote{Further investigation, beyond the work in this paper, is needed to give more insight into the selection of $K_{rel}$} (e.g., ACU). Therefore, the selection of $K_{rel}$ greatly depends on the requirements of a slice and state of the operator's network. The natural defense against malicious co-residency comes at no additional computational cost to the network operator since the cost is already included in the slice allocation.

\textbf{Acknowledgment:} The second author acknowledges funding from Canada’s NSERC through the Discovery Grant Program.
\squeezeup
\bibliographystyle{IEEEtran}
\bibliography{5Gbib}
\end{document}